\documentclass[conference]{IEEEtran}
\ifCLASSINFOpdf
  \usepackage[pdftex]{graphicx}
\else
\fi
\usepackage{balance}
\usepackage{wrapfig}

\usepackage{amsmath}
\usepackage{fancyhdr}
\usepackage{kantlipsum}
\fancypagestyle{plain}{
\fancyhf{}
\fancyhead[C]{Conference on \LaTeX} %% C or L or R.
}
\usepackage{eso-pic}

\begin{document}

% Code snippet 2: Just after the \begin{document} command insert the following code to add the conference header:
\AddToShipoutPictureBG*{
\AtPageUpperLeft{
\setlength\unitlength{1in}
\hspace*{\dimexpr0.5\paperwidth\relax}%% change \dimexpr0.5\paperwidth\relax appropriately
\makebox(0,-0.75)[c]{\textbf{2020 IEEE/ACM International Conference on Advances in Social
Networks Analysis and Mining (ASONAM)}}}}

\title{Understanding Public Sentiments, Opinions and Topics about COVID-19 using Twitter}

\author{\IEEEauthorblockN{Jolin Shaynn-Ly Kwan}
\IEEEauthorblockA{Information Systems Technology and Design Pillar\\
Singapore University of Technology and Design\\
Email: jolin\_kwan@mymail.sutd.edu.sg}
\and
\IEEEauthorblockN{Kwan Hui Lim}
\IEEEauthorblockA{Information Systems Technology and Design Pillar\\
Singapore University of Technology and Design\\
Email: kwanhui\_lim@sutd.edu.sg}}


%Telephone: (800) 555--1212, Fax: (888) 555--1212}


\maketitle

% for IEEE copyright statement
% Code snippet 3: Please insert the following code just after the /maketitle and before /begin{abstract} of the template in order to add the copyright strip
\IEEEoverridecommandlockouts
\IEEEpubid{\parbox{\columnwidth}{\vspace{8pt}
\makebox[\columnwidth][t]{IEEE/ACM ASONAM 2020, December 7-10, 2020}
\makebox[\columnwidth][t]{978-1-7281-1056-1/20/\$31.00~\copyright\space2020 IEEE} \hfill}
\hspace{\columnsep}\makebox[\columnwidth]{}}
\IEEEpubidadjcol

% for page number
%\thispagestyle{plain}
%\pagestyle{plain}

\begin{abstract}
The COVID-19 pandemic has caused widespread devastation throughout the world. In addition to the health and economical impacts, there is an enormous emotional toll associated with the constant stress of daily life with the numerous restrictions in place to combat the pandemic. To better understand the impact of COVID-19, we proposed a framework that utilizes public tweets to derive the sentiments, emotions and discussion topics of the general public in various regions and across multiple timeframes. Using this framework, we study and discuss various research questions relating to COVID-19, namely: (i) how sentiments/emotions change during the pandemic? (ii) how sentiments/emotions change in relation to global events? and (iii) what are the common topics discussed during the pandemic?
\end{abstract}


% \fi
\IEEEpeerreviewmaketitle

\section{Introduction}

The COVID-19 outbreak is a global pandemic that has infected millions and claimed the lives of hundred thousands. Social media, e.g., Twitter, has proved to be valuable in times of such global pandemics, as it reveals the real-time sentiment and discussions of Twitter users~\cite{Xue2020}. Additionally, with the implementation of quarantine measures and the prohibition of social gatherings across the world, the usage of social media has soared to extraordinary levels. 

Towards this effort, we aim to answer three research questions through analysis of global tweets about COVID-19 in the duration after the virus was named as a pandemic. The research questions (RQ) are:

\begin{itemize}
  \item RQ1: What are the sentiments and emotions experienced during the COVID-19 pandemic? 
  \item RQ2: How do these sentiments and emotions change in relation to global events?
  \item RQ3: What are the common topics discussed during the COVID-19 pandemic? 
\end{itemize}

{\bf Related Work}. Social media data have been a popular source of information for studying pandemics and diseases~\cite{Liu-BigData20}. For example, Signorini et al. studied the use of Twitter to track rapidly-evolving public interest with respect to the 2009 H1N1 pandemic in the United States~\cite{signorini2011}. The study found that interest in hand hygiene and face masks seemed to be timed with public health messages from the Centers for Disease Control and Prevention (CDC). Chew studied public sentiment with regards to the 2009 H1N1 pandemic from manual content analysis of tweets~\cite{chew2010}. Keywords corresponding to certain sentiments were collated and the tweets were categorised into the different sentiments by string detection of the different keywords.

Other researchers worked on different aspects of studying the COVID-19 pandemic. Dubey~\cite{dubey2020} studied the sentiments of COVID-19 related tweets between 11-31 March 2020, while Rajput et al.~\cite{Rajput2020} used word frequency analysis to identify trends in the words used. Similarly, Lwin et al.~\cite{lwin2020global} studied the prevalence of the four emotions of fear, anger, sadness, and joy in relation to the COVID-19 pandemic. Abd-Alrazaq et al.~\cite{Abd-Alrazaq2020} focused on understanding topics related to COVID-19 from 2 February to 15 March 2020 and~\cite{bisanzio2020} used mobility patterns within and out of Wuhan, China to understand COVID-19 cases.

{\bf Contribution}. While there has been significant research with regards to pandemics, the use of Twitter as a research tool for this is still relatively new and focus on one aspect of the research. In contrast, our work adopts a holistic framework that combines sentiment analysis, emotion detection, topic modelling, spatial/temporal analysis to understand the range of issues surrounding COVID-19. For example, we combine sentiment analysis with topic modelling to provide a deeper understanding about how individuals feel about specific topics, which can aid public health officials in developing strategies and policies.

\section{Research Framework}

In this section, we discuss the framework used to utilize public tweets for understanding sentiments, emotions and discussion topics related to COVID-19.

\begin{figure*}[th]
  \begin{minipage}{0.495\textwidth}
    \includegraphics[width=\linewidth, trim=0mm 20mm 0mm 37mm, clip]{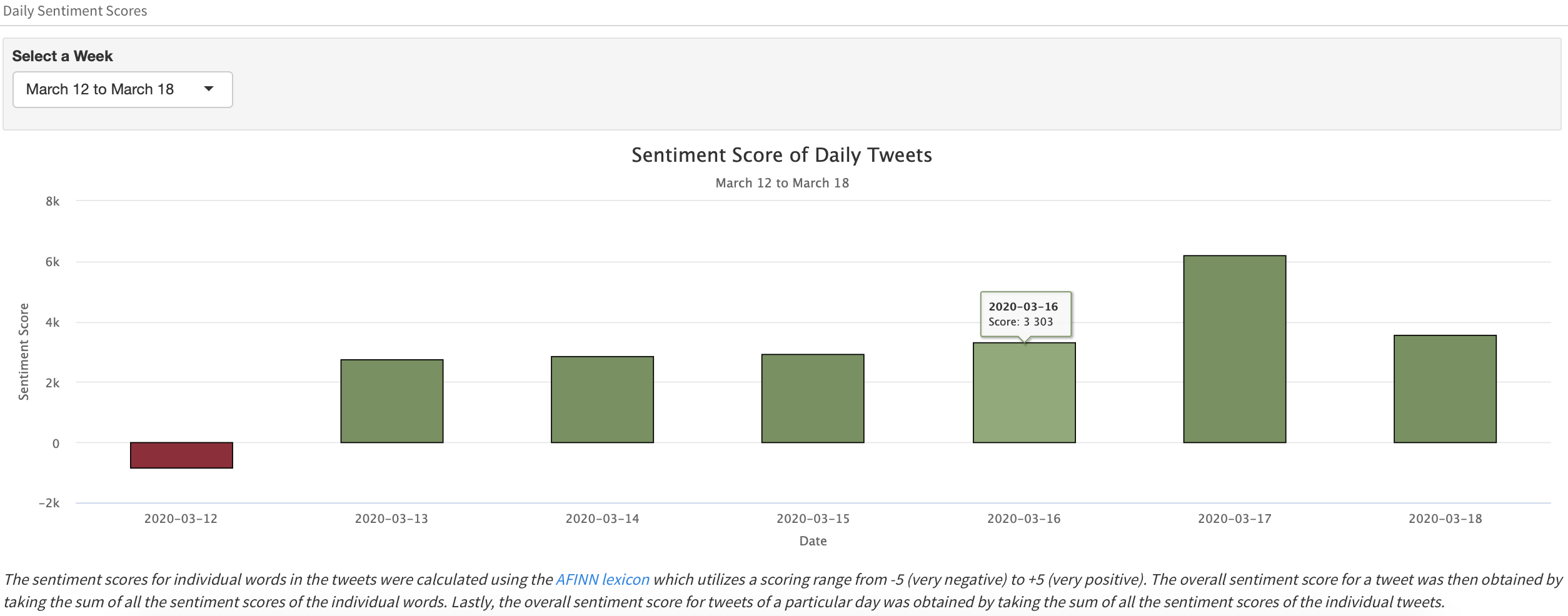}
    \caption{Time Series Plot for Sentiment Score of Daily Tweets}
    \label{figSentiment1}
  \end{minipage}
  \begin{minipage}{0.495\textwidth}
    \includegraphics[width=\linewidth, trim=0mm 20mm 0mm 38mm, clip]{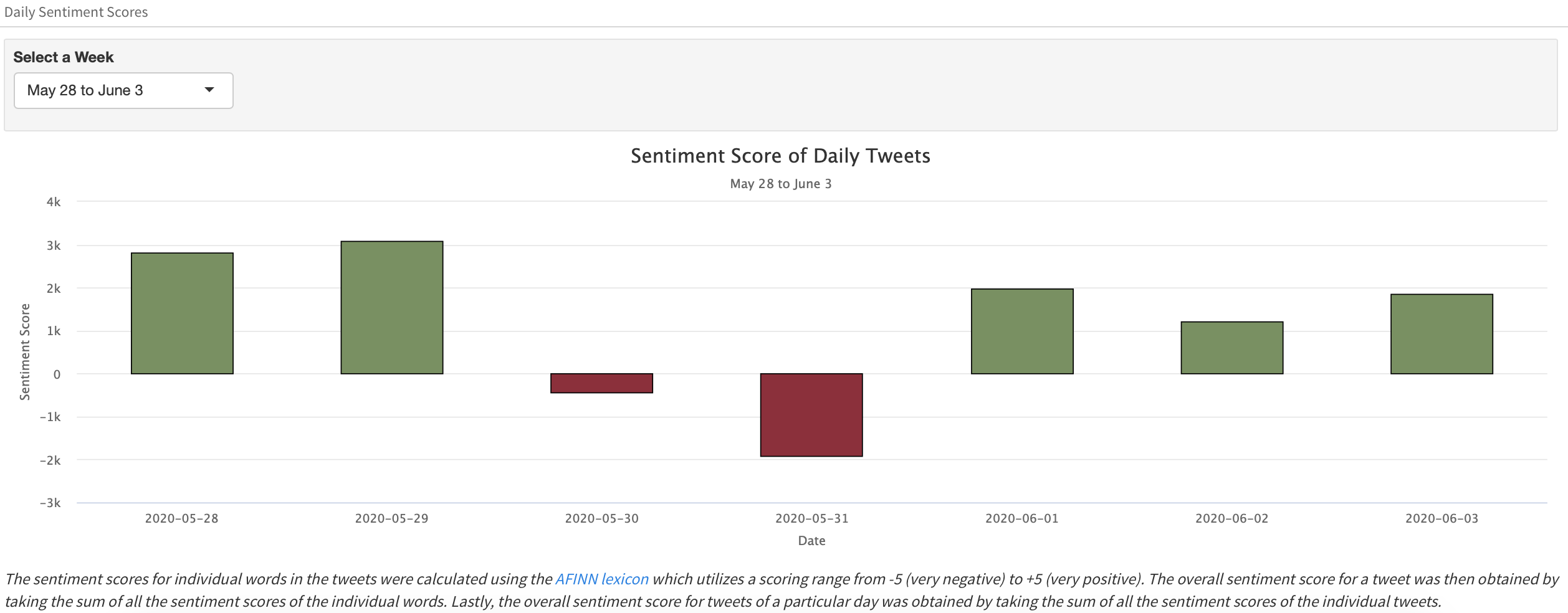}
    \caption{Sentiment Score of Daily Tweets for Week of 28 May to 3 June}
    \label{figSentiment2}
  \end{minipage}
\end{figure*}

\subsection{Data Collection} 
Our dataset comprises 910,000 public tweets from users across the world, curated from a combination of three larger COVID-19 related datasets. We extracted 70,000 tweets per week that were more likely to contain geolocation information (more details later), for the period of 12 March to 10 June 2020. Out of the 910,000 tweets, there are a total of 896,031 geolocated tweets in English. We extracted this subset of tweets from three larger datasets, namely: (i) Shane Smith’s datasets on Kaggle~\cite{Smith2020}, where daily datasets of tweets containing hashtags surrounding the topic of COVID-19 from 12 March to 30 April 2020; (ii) The Github repository of~\cite{Rrishik2020} contained tweet IDs of collected tweets from 1 May to 22 May 2020; and (iii) our own data collection from 23 May 2020 onwards.

\subsection{Data Preprocessing}
Previous studies have estimated that only 0.85\% of tweets are GPS geotagged~\cite{sloanMorgan2015,chi-wnut16}. 
Likewise for the original larger datasets, the proportion of tweets that are geotagged is extremely small. To address this issue, we utilized other attributes of the tweets to infer the location of these tweets, such as: (i) the self-declared location in the user profiles; and (ii) the “entities” section of the tweets which contain metadata and additional contextual information about them. This resulted in the significantly higher number of 896,031 geolocated tweets that will be used for spatial analysis. 
Thereafter, we perform the standard text preprocessing steps such as conversion to lowercase, tokenizing, lemmatizing and stemming.

\subsection{Sentiment/Emotion Analysis} 
The process of sentiment/emotion analysis involves taking the pool of words that have been identified from the preprocessing steps and comparing them to lexicons to determine their sentiments or polarity~\cite{lim-smartcity19}. There are two main lexicons used in this study: the NRC Word-Emotion Association Lexicon (also known as EmoLex) and the AFINN lexicon. The NRC lexicon is a list of more than 14,000 English unigrams and their associations with eight fundamental emotions (anger, fear, anticipation, trust, surprise, sadness, joy, and disgust) and two sentiments (negative and positive)~\cite{mohammadTurney2013}. The assignment and labelling of a word to its sentiment were manually done by crowdsourcing through Amazon’s Mechanical Turk service. The eight fundamental emotions in the lexicon correspond to the eight primary emotions introduced by Plutchik~\cite{plutchik1980}. 

The AFINN lexicon, on the other hand, assigns English words on a scale of -5 to 5 (excluding 0), with negative scores representing negative sentiment and positive scores representing positive sentiment~\cite{UniversityofCincinnati2018}. The sentiment score for each tweet is calculated as a sum of the sentiment scores for words in the tweet. A net positive overall sentiment score indicates a positive statement, a net negative overall sentiment score indicates a negative statement and a net overall sentiment score of zero indicates a neutral statement.

Therefore, two aspects of sentiment analysis were carried out for this research. The NRC lexicon was used for the detection of the different emotions that Twitter users have towards the pandemic, while the AFINN lexicon was used for detecting the extent of negativity/positivity of the tweets regarding the pandemic.

\subsection{Topic Modelling} 
Latent dirichlet allocation (LDA), a topic modelling algorithm, was carried out to identify key topics talked about in tweets surrounding the COVID-19 pandemic. LDA is a generative probabilistic model of a corpus~\cite{blei2003}. The algorithm works by representing documents as random mixtures over latent topics, where each topic is characterized by a collection of words (“bag-of-words”)~\cite{montenegro2018}.

The tweets were grouped into weeks (7 days) with a total of 13 weeks and the tweet pool of each individual week is treated as an individual document. To identify the optimal number of topics for LDA on the 13 different documents, the measure of perplexity was used. Perplexity is a statistical measure of how well a probability model predicts a given sample. In the context of LDA, the theoretical word distribution represented by the given number of topics is compared to the actual topic mixtures or distribution of words in the documents~\cite{UniversityofChicago2020}.

\begin{figure*}[th]
  \begin{minipage}{0.495\textwidth}
    \includegraphics[width=\linewidth, trim=0mm 0mm 0mm 39mm, clip]{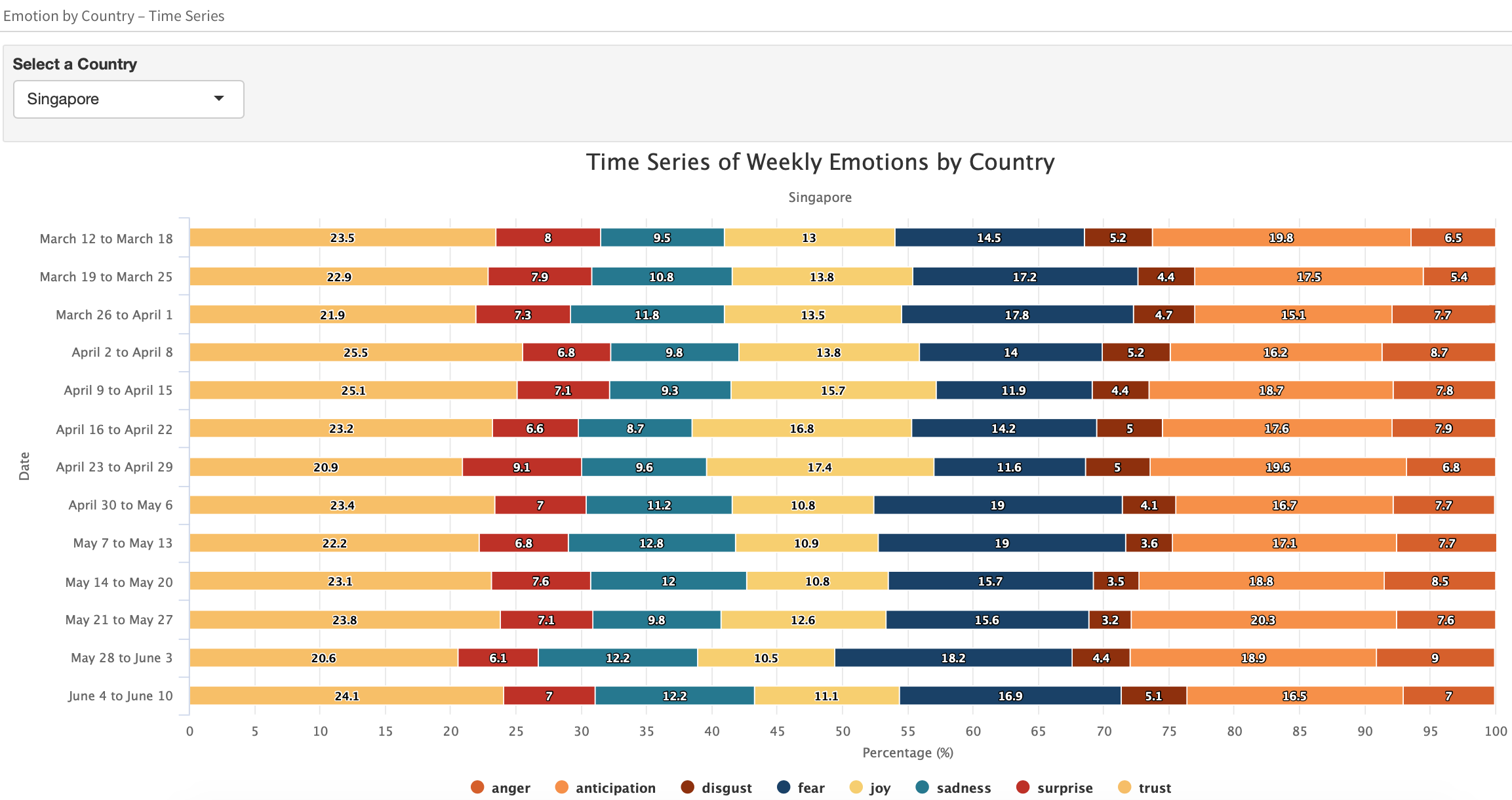}
    \caption{Time Series of Weekly Emotions in Singapore}
    \label{figEmoSg}
  \end{minipage}
  \begin{minipage}{0.495\textwidth}
    \includegraphics[width=\linewidth, trim=0mm 0mm 0mm 39mm, clip]{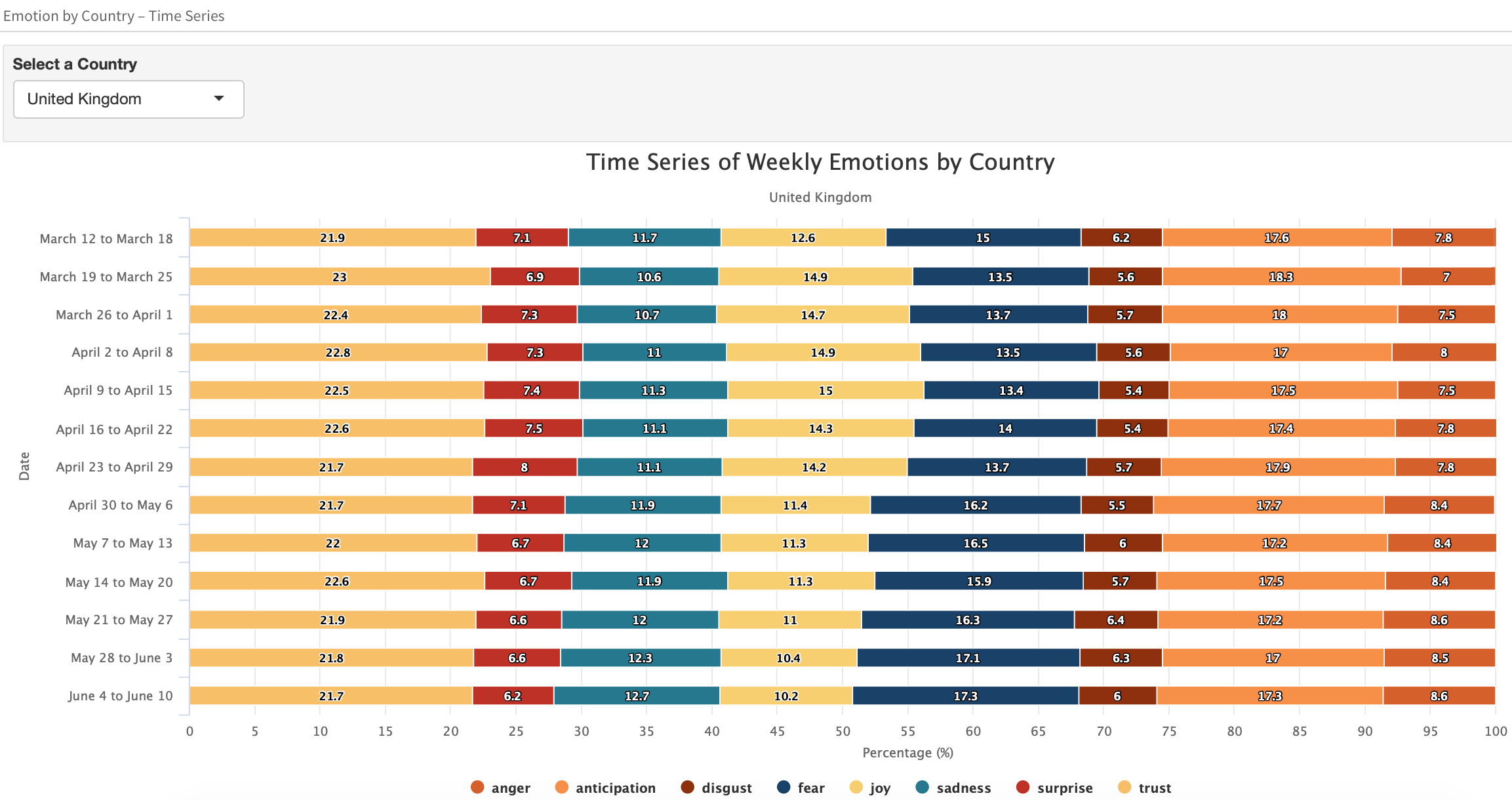}
    \caption{Time Series of Weekly Emotions in the United Kingdom}
    \label{figEmoUK}
  \end{minipage}
\end{figure*}

\section{Findings and Discussion}

\subsection{RQ1: Is it all doom and gloom?} 

We observe that the emotion “trust” surprisingly makes up a consistently higher proportion of tweets as compared to other emotions. Additionally, the emotion “disgust” makes up for a consistently lower proportion of tweets as compared to others. These results are surprising because a pandemic is usually associated with devastation and it would be natural to assume a greater prevalence of negative emotions. These negative emotions have been reported to stem from areas such as the fear and anxiety people feel in the face of a novel virus, the feelings of isolation and loneliness from social distancing~\cite{CDC2020}, as well as the stress and worry from economic uncertainties~\cite{Bareket-Bojmel2020}. However, a study by Barford et al. found that negative emotions rarely overwhelm positive ones in everyday circumstances~\cite{barford2020}, and the findings from our research displays the same phenomenon even in the midst of the COVID-19 pandemic. A study surveyed 854 Australian residents about their emotions specifically during the COVID-19 pandemic in March 2020 and the results showed that only 3\% of the participants felt purely negative emotions, but almost all the participants actually reported feeling positive emotions at some point in time. These positive emotions were attributed to factors such as the novelty and flexibility of different working arrangements and also feelings of approval towards social distancing for the sake of everyone’s collective health~\cite{Anglim2020}.

\subsection{RQ2: Global Events and Sentiments} 
    
The first global event that will be covered is the declaration of COVID-19 as a pandemic by WHO Director-General on 11 March 2020. This came after the number of cases COVID-19 outside China increased by 13 folds and the number of affected countries tripled in just two weeks~\cite{WHO2020}. This study started the analysis of tweets on 12 March, the day after the declaration and it can be seen from the visualization of daily sentiment score (see Figure~\ref{figSentiment1}) that people’s sentiments were significantly more negative on 12 March as compared to other days within the week as well as in the next few weeks. This is likely due to the fact that anxiety and fear levels were heightened after people realised the gravity of the COVID-19 situation and how many aspects of their lives will be impacted -- be it in daily routines or in finances.

The second global event is the protests sparked by the death of George Floyd on 25 May 2020 where a police officer killed him by kneeling on his neck during an arrest. The Black Lives Matter movement kick-started many protests not just in the USA, but all over the world as well. The weekend of May 30 and May 31 saw protests all over the world in countries like Britain, Germany, France and Spain, where people stood “in a global act of solidarity”~\cite{Colarassi2020}. To add on to the anger and devastation that people had towards this issue, the President of the USA, Donald Trump tweeted controversial and insensitive comments in response to the protests. His tweets on 29 May and 30 May threatened violence against the protesters outside the White House with one tweet saying that they could have been attacked with “vicious dogs and ominous weapons” by the US Secret Service~\cite{Milman2020}. The culmination of the protests and the comments by Trump fuelled people’s rage towards this situation and this can be seen in the significantly more negative sentiment scores for 30 May and 31 May (see Figure~\ref{figSentiment2}).

Figure~\ref{figEmoSg} shows the time series of the proportions of emotions exhibited by Twitter users in Singapore with regards to COVID-19. The week after the circuit breaker\footnote{Measures to reduce movement, activity and interactions in the country.} started for Singapore on 7 April (April 9 to April 15), it can be seen that percentages of the emotions “fear” (11.9\%) and “sadness” (9.3\%) dropped to one of the lowest percentages for their respective emotions throughout the whole duration of the dataset; while the percentages of the emotions “trust” (25.1\%) and “joy” (15.7\%) rose to one of the highest percentages for their respective emotions. From these results, it can be seen that contrary to popular belief that being in a “lockdown” dampens moods with the prohibition of social gatherings as well as harsh consequences associated with breaching the associated rules, Singaporeans generally still do feel a sense of relief and satisfaction at the implementation of the circuit breaker. According to a recent analysis of Singaporeans’ sentiment online, positive sentiments could be attributed to the Singapore government’s initiatives such as the Solidarity Budget that offers help to Singaporeans’ livelihoods while coping with the economic effects of the pandemic, and also the sense of relief at stricter control measures in light of growing cases of infection~\cite{Syed2020}.

Figure~\ref{figEmoUK} shows the time series of the proportions of emotions exhibited by Twitter users in the United Kingdom with regards to COVID-19. Looking at the percentages of the emotions across the weeks, it can be seen that the percentages for the emotions “fear” (16.2\%) and “sadness” (11.9\%) saw significant increases while emotions like “joy” saw a significant decrease (11.4\%) starting from the week of April 30 to May 6. During the first coronavirus press conference since being infected with COVID-19 on 30 April, the Prime Minister of the United Kingdom (UK) , Boris Johnson, announced that the UK was “past the peak” of the virus outbreak and that a “road map” would be laid out to phase out the lockdown in the UK~\cite{Triggle2020}. The “road map” was eventually revealed on 10 May where Johnson announced a phased end to existing lockdown restrictions. However, many criticisms were made against the way Johnson handled the reopening -- that there was a lack of clear plans and guidance for safety~\cite{BienkovColson2020} at a critical time in managing the pandemic~\cite{Forsyth2020}. In addition, the change of message for the UK’s coronavirus safety campaign from “stay home” to “stay alert” has also been criticized as “confused” and “nonsensical” by the public and rejected by nations such as Scotland and Wales~\cite{Handley2020}. The feelings of uncertainty and anxiety felt by people in the UK during this period is reflected in how the overall levels of negative emotions increased significantly from the week of April 30 onwards, while the overall levels of positive emotions decreased.

\subsection{RQ3: Discussion Topics} 

The analysis from longitudinal topic modelling for 3 months of tweets related to COVID-19 has proven to be effective in identifying key trends of public dialogue and could potentially be used to guide targeted and timely interventions. For example, the overall trend of topics started off containing words with negative connotations such as “violence”, “assault”, “marginalize”, with the topics suggesting the woes and frustrations associated with the COVID-19 lockdown/quarantine. However, as time went by, the topics started shifting towards the more positive end of the spectrum, with topics that involved tweets about keeping productive during the lockdown and tweets about approaching a “new normal” with telecommuting and e-commerce. This signals a potential “glimmer of hope” amidst the bleak outlook because it shows that society is accelerating the use of technology, progressing forward in becoming “smart” cities, and adapting to the new future.

\section{Conclusion} 

In this paper, we present a framework that allows the general public or public health officials, to understand the public sentiment towards the pandemic, the key topics discussed. For example, with an idea of what the key topics are at a specific time in point, as well as across time, the general public can be educated about events that might not have been reported that widely in their specific countries. Using this framework, we studied three research questions relating to COVID-19, namely: (i) how sentiments/emotions change during the pandemic? (ii) how sentiments/emotions change in relation to global events? and (iii) what are the common topics discussed during the pandemic?. We believe that this research’s methodology can be easily tweaked and applied to different contexts as well to examine other research topics such as that of a natural disaster or political events.

%\section*{Acknowledgment}
\vspace{3mm}
{\noindent\bf Acknowledgements}.
This research is funded in part by the Singapore University of Technology and Design under grant SRG-ISTD-2018-140. The authors would like to thank Olivia Nicol for her useful suggestions.

\bibliographystyle{IEEEtran}
\balance
\bibliography{covidTwitter}

\end{document}